\documentclass[aps,jstat,twocolumn]{revtex4}
\usepackage{graphicx}
\usepackage{amsmath}
\usepackage{braket}
\usepackage{hyperref}
\usepackage{subfigure}
\usepackage[usenames,dvipsnames]{color}

\usepackage{graphicx}
\usepackage{subfigure}

\newcommand{\ct}{\cite}

\newcommand{\bi}{\bibitem}
\newcommand{\be}{\begin{equation}}
\newcommand{\ee}{\end{equation}}
\newcommand{\ba}{\begin{eqnarray}}
\newcommand{\ea}{\end{eqnarray}}

\newcommand{\ga}{\gamma}
\newcommand{\non}{\nonumber}

\begin{document}
\title{Dynamical merging of Dirac points  in the periodically driven Kitaev honeycomb model}
\author{Utso Bhattacharya$^{1}$, Sayak Dasgupta$^2$ and  Amit Dutta$^1$\\
$^1$ Department of Physics, Indian Institute of Technology, 208016, Kanpur\\
$^2$ Department of Physics and Astronomy,
The Johns Hopkins University, 3400 N. Charles Street, Baltimore, MD 21218 }

\begin{abstract}
We study the effect of a half wave rectified sinusoidal electromagnetic (EM) wave on the Kitaev honeycomb model with an additional magneto-electric coupling term {arising due to induced polarization of the bonds. Within the framework of Floquet analysis, we show that merging of a pair of Dirac points in the gapless region of the Kitaev model leading to a semi-Dirac
spectrum is indeed possible}  by externally varying the amplitude and the phase of the EM field.  
\end{abstract}
\maketitle
\section{Introduction}
\label{sec_intro}
The Kitaev honeycomb model \ct{kitaev06} is an exactly solvable anisotropic spin-$\frac{1}{2}$ model with Ising-like interactions $\sigma^{x}_{r}\sigma^{x}_{r'}$, $\sigma^{y}_{r}\sigma^{y}_{r'}$ and $\sigma^{z}_{r}\sigma^{z}_{r'}$ assigned to the three bonds in the honeycomb lattice; 
{$\sigma^{\alpha}_{r/r'}$ stand for the $\alpha(=x,y,z)$-th Pauli spin operator residing on the $r/r'$-th site of the lattice}.   {Although there has not been an experimental realization of the Kitaev honeycomb spin model till date,  there exist some promising proposals of experimental realization of the spin models involving ultracold atoms \ct{duan00}  and polar molecules \cite{micheli06} in optical lattices. There have also  been theoretical proposals that the Kitaev model may be realized as an effective low-energy model for a Mott insulator with strong spin-orbit coupling \ct{jackeli09,chaloupka10}. 
}The model exhibits a rich phase diagram where depending on the relative strength of coupling  between the Ising interactions the system either has a graphene \cite{neto09} like gapless spectrum or a gapped spectrum; {the gapless phase and the gapped phases are separated by the phase boundaries  where the spectrum  is of semi-Dirac nature
 \ct{banerjee09,patel12} such that the associated quantum critical point \cite{sachdev99,suzuki13} is of an anisotropic nature \ct{dutta15}.

In the gapless phase of the model the dipersion relation $\epsilon(\vec{k})$ vanishes at the contact points between two bands. At these so called Dirac points the spectrum is linear and due to time reversal symmetry of the Hamiltonian these points occur in pairs in the reciprocal space at points $\vec{D}$ and $-\vec{D}$. It has been observed that for  graphene-like systems varying the hopping parameters, by application of periodic perturbations, leads to a shift in the point location and may lead to merging of two such Dirac points. On merger the resulting spectrum is a semi-Dirac spectrum with a dispersion relation which is quadratic along one direction in the reciprocal space and linear along the other (perpendicular) direction. The merging of Dirac points is accompanied by a topological phase transition from a semimetallic to an insulating phase \ct{delplace11,kim12,koghee12, delplace13,carpentier13}. We note that the magnetic field dependence of Landau levels in a Graphene-like structure resulting in a semi-Dirac spectrum was already  reported much earlier \ct{dietl08,mont09,mont091}.

{It is worthwhile to mention here that the study of topological insulators and the topological protection of edge states\ct{inoue10,kitagawa10,lindner11,
jiang11,trif12,gomez12,dora12,cayssol13,liu13,tong13,rudner13,katan13,lindner13,
kundu13,basti13,schmidt13,reynoso13,wu13,manisha13,perez14,reichl14,manisha14,kitagawa12,rechtsman13,puentes14,gu11} have gained a huge importance in recent years. In parallel, an extensive amount of work have also been dedicated to study in depth, driven closed quantum systems from the perspective of dynamical saturation\ct{russomanno12}, dynamical localization\ct{alessio13,bukov14,nag14,nag15}, dynamical freezing\ct{das10}, dynamical fidelity\ct{sharma14}, defect production\ct{mukherjee08,mukherjee09} and thermalization\ct{lazarides14}. But it is the advent of irradiated (Floquet) grapehene\ct{gu11,kitagawa11,morell12} and Floquet topological insulators that have gelled the two, facilitating experimentalists to be able to probe and verify a few of the many recent theoretical work done on the effect of driving on closed topological quantum systems\ct{kitagawa12,rechtsman13,puentes14}.}

Recently, a large body of work has also been done in which a time-periodic perturbation has been used to dynamically change the phases of Hamiltonians exhibiting {different} topological properties\ct{delplace13,inoue10,lindner11,gu11,kitagawa11,gomezleon13,cayssol13,thakurathi13,thakurathi14,rajak14}. In the context of Graphene, Delplace $et~al$ \ct{delplace13} used an external time-periodic electromagnetic (EM) field to merge a pair of Dirac points in the high frequency regime by varying the amplitude and the phase of the field applied. {On the other hand}, in our work we look at the equivalent problem in the Kitaev model. However, {there is a subtle issue that needs to be addressed};  the Kitaev model unlike graphene consists of {localized spins} which do not possess linear momenta  and hence the minimal coupling scheme used in [\onlinecite{delplace13}] cannot be extended to our case. To overcome this situation, we employ the scheme introduced by Sato $et~al$ \ct{sato14}  where the effect of an external EM field on the Kitaev honeycomb model was studied  by incorporating a magneto-electric coupling term 
 in the Hamiltonian. We use this modified Hamiltonian to study the possibility of merging the Dirac points in the gapless phase of the {original Kitaev model} through an external eliptically polarized, half wave rectified electromagnetic wave. Such a manipulation allows us to control the phases and hence the topological properties of the Kitaev model {driven by an  external periodic perturbation in the high-frequency limit}.\\   
 
The main motivation of our work is to externally tune the couplings of the Kitaev Hamiltonian in such a way so that the Dirac points in the gapless region merge leading to a semi-Dirac spectrum within the gapless region of the undriven Kitaev model itself. In our subsequent discussion we will present the fact that unlike Graphene with nearest neighbor hopping, the linear dispersion (or the Dirac point) in the gapless region of the undriven Kitaev model is anisotropic (with different velocities in the two different directions). Moreover, the presence of localized spins on lattice sites brings forth further difficulty while trying to couple the spins to any perturbation associated with an external electric field, in stark contrast to the case of Graphene, where the electron momentum can easily couple with the applied electromagnetic field via Peierls' substitution. Hence, in spite of all such difficult issues how merging of Dirac points in the gapless region of the undriven Kitaev model can still be achieved via a periodic external electromagnetic field, is an intriguing question of non-equilibrium statistical mechanics. The fact that the Kitaev model is the only exactly solvable model in two dimensions
provides us with further motivation.

The rest of the paper is organized as follows; the Kitaev model is outlined in Sec.\ref{sec_Kitaev}. {We describe the modifications necessary to make  the  Hamiltonian  couple to  an EM wave in Sec.\ref{sec.light_Kitaev}}. This is followed by a review of Floquet theory in Sec.\ref{sec_Floq} {generalizing it to the context of} the space periodic Kitaev model in Sec.\ref{sec_floq_bloch_kit}. We discuss the effect of driving on the Hamiltonian in Sec.\ref{sec_merge} and the resulting phase diagram in Sec.\ref{sec_phase}.  {Concluding comments are presented in Sec.\ref{sec_conc}.}

\section{Kitaev Model}
\label{sec_Kitaev}
The Kitaev model consists of spin-1/2's placed on the sites of a honeycomb lattice with a Hamiltonian of the form:
\begin{eqnarray}
\label{eq.Kit_ham}
H_{\rm Kitaev} &=& \sum_{j+l = even}(J_{1}\sigma^{1}_{j,l}\sigma^{1}_{j+1,l} + J_{2}\sigma^{2}_{j-1,l}\sigma^{2}_{j,l} \non\\
&+& J_{3}\sigma^{3}_{j,l}\sigma^{3}_{j,l+1})
\end{eqnarray}
where $j,l$ are the column and row indices respectively, $\sigma^{\alpha}_{m,n}$ are Pauli matrices, representing the spins, at the site labeled $(m,n)$ and $J_{i}$'s are the corresponding coupling parameters (see Fig.~\ref{fig_kitaev_model}).
In this section the energy spectrum for the time independent Kitaev model {and the corresponding phase  will be reviewed before we
proceed to discuss  how the addition of an EM coupling between the spins and an external field changes the spectrum by affecting the interaction strengths ($J_{i}$'s)}. We set $J_{i}\geq 0$  below without any loss of generality.\\
\begin{figure}
\includegraphics[height= 6cm, width=6cm]{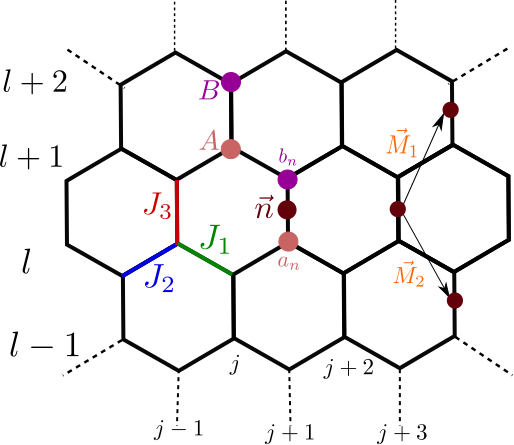}
\caption{(color online) This shows the honeycomb lattice of the Kitaev model with the interactions labeled as $J_{i}$. The vectors $\vec{M_{1,2}}$ span the lattice, and the positions of the sublattice points are labeled by $A$ and $B$.}
\label{fig_kitaev_model}
\end{figure}

{Referring to  Fig.\ref{fig_kitaev_model}}
we take the unit cells of the lattice to be the vertical bonds with sites labeled A and B,  respectively 
{so that there are  $N/2$ unit cells for a system of $N$ sites.}  {Choosing the nearest neighbor distance to be $1/\sqrt{3}$ we  label each unit cell} by a vector $\vec{n} = \widehat{i}n_{1} + \left(\frac{1}{2}\widehat{i} + \frac{\sqrt{3}}{2}\widehat{j}\right)n_{2}$, with $n_{1,2}$ {denoting} the coordinates of the B site in the unit cell $\vec n$. The spanning vectors   {joining} the neighbouring unit cells are then given by $\vec{M}_{1}=\frac{1}{2}\widehat{i} + \frac{\sqrt{3}}{2}\widehat{j}$ and $\vec{M}_{2} = \frac{1}{2}\widehat{i} - \frac{\sqrt{3}}{2}\widehat{j}$.
To fermionize the system we define the Majorana operators as:
\begin{eqnarray}
\label{eq.maj_op}
a_{j,l} \equiv \left(\prod^{j-1}_{i=-\infty}\sigma_{i,l}^{3}\right)\sigma_{j,l}^{2}, \hspace*{3 mm} {\rm for} \hspace*{1mm} j+l = {\rm even} \\ \nonumber
a'_{j,l} \equiv \left(\prod^{j-1}_{i=-\infty}\sigma_{i,l}^{3}\right)\sigma_{j,l}^{1}, \hspace*{3 mm} {\rm for} \hspace*{1mm} j+l = {\rm even} \\ \nonumber
b_{j,l} \equiv \left(\prod^{j-1}_{i=-\infty}\sigma_{i,l}^{3}\right)\sigma_{j,l}^{1}, \hspace*{3 mm} {\rm for} \hspace*{1mm} j+l = {\rm odd} \\ \nonumber
b'_{j,l} \equiv \left(\prod^{j-1}_{i=-\infty}\sigma_{i,l}^{3}\right)\sigma_{j,l}^{2}, \hspace*{3 mm} {\rm for} \hspace*{1mm} j+l = {\rm odd}
\end{eqnarray}
These are Hermitian operators {representing real fermions (i.e., $a^{\dagger} = a$)}  satisfying the anti-commutation relations {for example}, $\left\lbrace a_{m,n},a_{m',n}\right\rbrace = 2\delta_{m,m'}\delta_{n,n'}$.
  In terms of these operators the Hamiltonian (\ref{eq.Kit_ham}) takes the form:
\begin{equation}
\label{eq.Kit_ham_maj}
H = i \sum_{\vec{n}}(J_{1}b_{\vec{n}}a_{\vec{n}-\vec{M_{1}}} + J_{2}b_{\vec{n}}a_{\vec{n}-\vec{M_{2}}} + J_{3}D_{\vec{n}}b_{\vec{n}}a_{\vec{n}}).
\end{equation}
{Remarkably, operators $D_{\vec{n}} = b'_{\vec n} a'_{\vec n}$'s  defined on each plaquette
 commute with each other and with the Hamiltonian and their eigenvalues can take the values $\pm 1$ independently for each $\vec{n}$ This enables us to reduce  $2^{N}$-dimensional Hilbert space into $2^{N/2}$ dimensions. However, it has been established} that the global ground state of the model lies in the sector in which $D_{\vec{n}} = 1$ for all $\vec{n}$. To extract the spectrum we move into the Fourier space summing over half the Brillouin zone {due to the Majorana nature of the fermions}. The Hamiltonian can then be written in the $(2\times 2)$
Bogoliubov-de Gennes form as:
\begin{eqnarray}
\label{eq.Kit_ham_bdg}
H_{\vec{k}} =&& 2\left[J_{1}\sin(\vec{k}.\vec{M_{1}}) - J_{2}\sin(\vec{k}.\vec{M_{2}})\right]\tau^{x} \\ \nonumber
&&+2\left[J_{3} + J_{1}\cos(\vec{k}.\vec{M_{1}}) + J_{2}\cos(\vec{k}.\vec{M_{2}})\right]\tau^{y}
\end{eqnarray}
where $\tau^{\alpha}$ are the Pauli matrices denoting the pseudospins. {The dispersion relation can then
be immediately derived as}
\begin{eqnarray}
\label{eq.kit_spec}
E_{\vec{k}}^{2} = && 4\left(J_{1}\sin(\vec{k}.\vec{M_{1}}) - J_{2}\sin(\vec{k}.\vec{M_{2}})\right)^{2} \\ \nonumber
&& + 4\left(J_{3} + J_{1}\cos(\vec{k}.\vec{M_{1}}) + J_{2}\cos(\vec{k}.\vec{M_{2}})\right)^{2}.
\end{eqnarray}
The phase diagram of the model as deduced from Eq.~(\ref{eq.kit_spec}) is shown in  Fig.~\ref{fig_kitaev_phases}. Given that $J_{i}\geq 0$, it is convenient to choose a normalization $\sum_{i} J_{i} = 1$ which describes points lying within (or on) an equilateral triangle. This triangle can be divided into four smaller equilateral triangles. They are labeled as $A_{1}$ where $J_{1} > J_{2} + J_{3}$, $A_{2}$ where $J_{2} > J_{1} + J_{3}$ and $A_{3}$ where $J_{3} > J_{1} + J_{2}$. In the region  $B$ each of the $J_{i}$ is less than the sum of the other two. It turns out the spectrum is gapped in all the $A$ regions, and gapless in the $B$ region .\\

{We reiterate that the purpose of our work is to tune the couplings of the Kitaev Hamiltonian given in Eq.~(\ref{eq.Kit_ham}) through an external control, such that   Dirac points in the $B$ region merge leading to a semi-Dirac spectrum even within the
gapless phase of the unperturbed Kitaev model. This  we achieve using a periodic external electromagnetic field where the  phase and amplitude of the field renormalizes the hopping strengths $J_{i}$'s leading to the desired merging.}
\begin{figure}
\includegraphics[height = 5cm, width = 7cm]{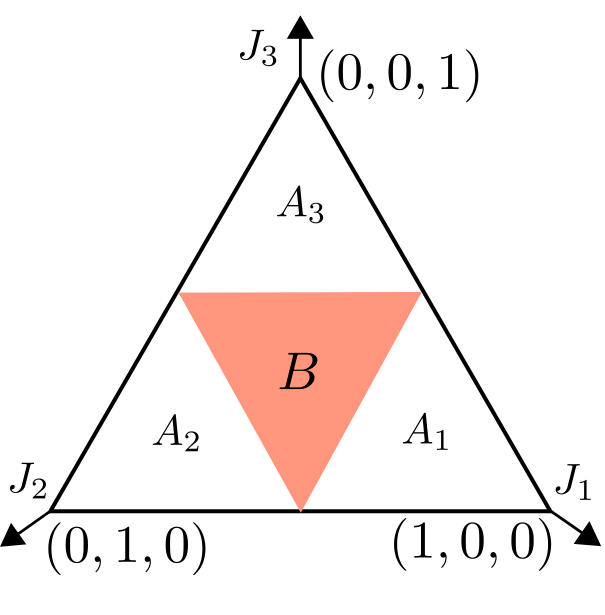}
\caption{(color online) This shows the phase diagram of the Kitaev model. The plane $J_1+J_2+J_3=1$ in the parameter space described by $(J_1,J_2,J_3)$ has been chosen for illustration without any loss of generality. The phases marked $A_{1,2,3}$ are gapped while the phase marked $B$ is gapless. We highlight the two phases (A and B), that we have concentrated on, in the discussion.}
\label{fig_kitaev_phases}
\end{figure}

\section{Light Impingement on the Kitaev model}
\label{sec.light_Kitaev}
{As discussed above that} the Kitaev model is essentially a system of {localized} spins on a lattice {without any   real momentum which could couple to the EM field as happens in the case of itinerant  electrons in  Graphene}. {To surmount this difficulty,  Sato $et~al$  \ct{sato14}  introduced a  magneto-electric coupling between the EM field and the spin lattice}.\\

In our approach, it is assumed that the EM field affects the underlying lattice by causing
the stretching (or compression) of the underlying lattice bonds those host
the spins; therefore, the EM fields leads to an effective polarization  
enabling us to define an effective polarization tensor  that couples to the external
field; this coupling  consequently  alters the bare bond strengths
$(J_{1},J_{2},J_{3})$ renormalizing them to $(J'_{1},J'_{2},J'_{3})$. For clarity, we would like to stress that the co-ordinate
axes $1$, $2$ and $3$, are oriented in the bond-direction in the spin space while $(x,y,z)$ represent spatial
cartesian coordinates.
The ME coupling  originates from electric  polarization on each bond and its strength {is assumed to be} proportional to the exchange interaction of the bond \ct{sato14}.
{The Hamiltonian now includes an additional piece}
\begin{equation}
\label{eq.ME_Kit_ham}
H(t) = H_{{\rm Kitaev}} + H_{\rm {ME}}
\end{equation}
{One assumes} a polarization of the form $\vec{P}_{(r,r')_{\alpha}} = \vec{\Pi}_{\alpha}\sigma^{\alpha}_{r}\sigma^{\alpha}_{r'}$, defined on each bond $(r,r')_{\alpha}$, where $\vec{\Pi}_{\alpha}$ is the ME coupling vector and $\alpha=1,2,3$ denote the 
direction of the bonds. This {leads to} a total polarization as ${\vec P}_{\rm total} = \sum_{\alpha}\sum_{(r,r')_{\alpha}}\vec{P}_{(r,r')_{\alpha}}$ so that $H_{\rm {ME}}(t) = -\vec{E}(t) \cdot {\vec P}_{\rm total}$. We have taken the polarization along the bonds of the Kitaev model
which lie in a plane of the Honeycomb lattice which in the
Cartesian basis $(x,y)$ is represented as:

\be \vec \Pi = \Pi_3 {\hat y} + \frac{1}{2} \Pi_1 (\hat x -\sqrt{3} \hat y)+\frac{1}{2} \Pi_2 (-\hat x -\sqrt{3} \hat y). 
\label{eq_Pi}
\ee

%
{In the subsequent discussion,  we shall work with} a Hamiltonian of the form:
\begin{equation}
\label{eq.Kit_ham_driven}
H = \sum_{j+l = even}\left(J'_{1}\sigma^{1}_{j,l}\sigma^{1}_{j+1,l} + J'_{2}\sigma^{2}_{j-1,l}\sigma^{2}_{j,l} + J'_{3}\sigma^{3}_{j,l}\sigma^{3}_{j,l+1} \right)
\end{equation}
using an electric field of the form $\vec{E}(t) =  A(\sin(\Omega t), \sin(\Omega t + \phi),0))$, where {the
renormalized coupling strengths are } $J'_{\alpha} \equiv (J_{\alpha} - \vec{\Pi}_{\alpha}.\vec{E})$ and the index $\alpha$ runs over the bonds $(1,2,3)$. 

  {Our work does rely on the magneto-electric (ME) coupling term (first introduced in Ref. \cite{sato14} in the case of the Kitaev model) to produce a merging of the Dirac points by light impingement; we would, however, like to emphasize  that  such an interaction term is not uncommon in multiferroic materials (which exhibit both ferroelectric and magnetic ordering) and has been experimentally detected in rare earth metal (R = Tb, Gd) perovskite manganites [RMnO$_3$] \cite{kimura03}. RMnO$_3$ are hexagonal polar crystals with ferrimagnetic, ferromagnetic or anti-ferromagnetic orders \cite{pimenov06}) and may be implemented in optical lattices as proposed  in Ref. \cite{micheli06} to generate the Kitaev spin model. The ME coupling is strong in such multiferroic materials due to the presence of frustrated magnetic exchange interactions and non-collinear spin order. Theoretically, the ME coupling term can easily be understood (in terms of the phenomenological theory developed by Landau) by expanding the free energy as a function of the magnetization and the electric polarization. As a result the linear ME effect arises from the cross term $\beta_{\mu\nu}M_\mu P_\nu$ where $\beta_{\mu\nu}$ is a second ranked tensor, $M_\mu$ and $P_\nu$ are the magnetization and the electric polarization vectors, respectively. This coupling also depends on the symmetry of the crystal and is only present when both the time reversal symmetry  and the inversion symmetry  are  broken. We recall that the  breaking of time reversal symmetry and the inversion symmetry  are  necessary for spontaneous magnetization and ferro-electric polarization, respectively. We refer to  Tokura et al. \cite{tokura14} for a discussion of the exchange-striction type of ME coupling used in our paper (in particular in the discussion below  Eq. [6]) and of the ME coupling in multiferroics.  In our model, we have assumed that such a coupling originates from a slight distortion of the underlying lattice due to the electric field of the laser light or through  a tiny phonon-mediated dimerization leading to a polarization of the bonds whose strength as a result is some fraction of the strength of spin-spin exchange interaction along the bonds.}

\section{Floquet-Bloch Theory in periodically driven systems:}
\label{sec_Floq}
The Floquet technique \cite{shirley65,griffoni88,stockmann99} (which is a temporal version of Bloch's theorem) deals with Hamiltonians subjected to a time-periodic potential of the form $H = H_{0} + V(t)$, where $V(t+\tau)=V(t)$. {Recalling the discrete time translation operator $T$ defined through the relation $T\psi_{\alpha}(x,t)=\psi_{\alpha}(x,t+\tau)=\lambda_{\alpha}\psi_{\alpha}(x,t)$,
we note that for stationary solution, $\lambda_{\alpha}$} has to be a pure phase of the form $e^{-i\phi_{\alpha}}$. Thus we have a solution of the form :
\begin{equation}
\label{eq.floq1}
\psi_{\alpha}(x,t+\tau) = e^{-i\omega_{\alpha}t}u_{\alpha}(x,t),
\end{equation}
{where $u_{\alpha}(x,t+\tau)$ is a time-periodic function that satisfies the condition $u_{\alpha}(x,t+\tau)=u_{\alpha}(x,t)$} and $\omega_{\alpha} = \phi_{\alpha}/\tau$. {In a spirit similar to Bloch theorem, where one defines
quasi momenta,  in the time periodic case one introduces the notion of  quasi energies} of the form $E_{\alpha} = \hbar \omega_{\alpha}$, defined within the first Brillouin zone, $(-\frac{\hbar}{2\tau},\frac{\hbar}{2\tau})$.  {When viewed
stroboscopically (at the end of each complete  period $\tau$), the problem effectively reduces to a time-independent problem; the
 unitary time evolution operator  after $n$-complete periods assumes a simple form given by $U(n\tau,0)={\cal T}\exp \left(-i\int_0^{n\tau} H(t)dt \right)=[U(\tau,0)]^{n}$, where ${\cal T}$ denotes the time ordering operator. Working in a representation in which  $U$ is diagonal such that $U_{D} = diagonal[e^{-i\phi_{n}}]$, one
 can evaluate the quasi energies in a straightforward manner}. \\

In {the present} case we are dealing with a Hamiltonian that is periodic in both space and time {and is} represented by:
\begin{equation}
\label{eq.gen_form_floq_bloc}
H(\vec{x} + \vec{a}_{i}, t + \tau) = H(\vec{x}, t),
\end{equation}
where $\vec{a}_{i}$'s are the lattice vectors. The Floquet-Bloch Hamiltonian is then given as:
\begin{eqnarray}
\label{eq.Ham_floq_bloch}
\non
\mathcal{H}(\tau,\vec{k}) &&= e^{-i\vec{k}.\vec{x}}H(\tau)e^{i\vec{k}.\vec{x}}-i\partial_{\tau} \\ 
\mathcal{H}(\tau,\vec{k}) &&= H_{\vec{k}}(\tau) - i\partial_{\tau}
\end{eqnarray}
where $H_{\vec{k}}(\tau)$ is in the $\vec{k}$-space.
The  Floquet states, defined earlier,  now get generalized  to Floquet-Bloch states which are periodic in both space and time and obey the equation $(H_{\vec{k}}(\tau) - i\partial_{\tau})\ket{u_{\alpha,\vec{k}}} = \epsilon_{\alpha,\vec{k}}\ket{u_{\alpha,\vec{k}}}$; where $\ket{u_{\alpha,\vec{k}}}$ is periodic in both space and time. The ket is defined in a composed Hilbert space (the Sambe space) which is the direct product space of the original Hilbert space and and {the space} of time($\tau$)-periodic functions \ct{sambe73}. The inner product in this space is defined by a composed scalar product:
\begin{equation}
\label{eq.inner_prod}
<< ... >>  = \frac{1}{\tau}\int_{0}^{\tau}<...>dt
\end{equation}
where $<...>$ is the normal inner product in Hilbert space.

{The above discussion leads to the conclusion} that using the Floquet-Bloch ansatz, one can reduce  both $\vec{k}$ and time to parameters in a time independent eigenvalue equation {although the dimensionality  of the Hilbert space gets augmented}. This {reduction} enables us to {exploit the nature of the} quasi-energy spectrum to study the properties of a driven Kitaev chain. Here, we have  multiple copies of the same Bloch bands in frequency space, {known as Floquet Bloch bands, while  the coupling between the copies is determined by the frequency of driving}. {We shall work in the  high frequency limit where there is no coupling between the different frequency Brillouin zones  restricting the system to the original Bloch band with renormalized hopping parameters.}
\section{Floquet-Bloch Hamiltonian for the driven Kitaev Model}
\label{sec_floq_bloch_kit}
In the case of the driven system, given in Eq.~(\ref{eq.Kit_ham_driven}), the couplings are time dependent. Consequently, to diagonalize this system we will need to introduce time dependent Majorana operators ($a,b$) which lead to the form

\begin{eqnarray}
\label{eq.drive_kit_jordan}
\non
H = &&i \sum_{\vec{n}}J'_{1}(t)b_{\vec{n}}(t)a_{\vec{n}-\vec{M_{1}}}(t) + J'_{2}b_{\vec{n}}(t)a_{\vec{n}-\vec{M_{2}}}(t) \\ 
 &&+ J'_{3}D_{\vec{n}}(t)b_{\vec{n}}(t)a_{\vec{n}}(t). 
\end{eqnarray}
Using the time dependent form of Eq.~(\ref{eq.Kit_ham_bdg}) in Majorana representation with the notation $a \to c_{a}$ and $b \to c_{b}$, we obtain a time dependent $(2\times 2)$ Hamiltonian of the form:
\begin{equation}
\label{eq.maj_ham_timed}
H_{\vec{k}}(t) = \sum_{\alpha,\beta} c^{\dagger}_{\alpha,\vec{k}}(t)h^{\alpha,\beta}_{\vec{k}}(t)c_{\beta,\vec{k}}(t)
\end{equation}
with both $\alpha$, $\beta$ running over the sublattice index $(a,b)$. The Hamiltonian matrix is given as:
\begin{equation}
\label{eq.kit_timed_bdg_maj}
H = \left(\begin{array}{cc} 0 & \rho(\vec{k},t)  \\ \cr \rho^{*}(\vec{k},t) & 0 \end{array} \right)
\end{equation}
with $\rho(\vec{k},t) = A_{\vec{k}}(t)-iB_{\vec{k}}(t)$ and 
\begin{eqnarray}
\label{eq.timed.breakup.ham}
A_{\vec{k}} &&= 2\left[J'_{1}(t)\sin(\vec{k}.\vec{M_{1}}) - J'_{2}(t)\sin(\vec{k}.\vec{M_{2}})\right] \\ \non
B_{\vec{k}} &&= 2\left[J'_{3}(t) + J'_{1}(t)\cos(\vec{k}.\vec{M_{1}}) + J'_{2}(t)\cos(\vec{k}.\vec{M_{2}})\right]
\end{eqnarray}
As the system is time-periodic the Majorana operators can be decomposed into their frequency Fourier modes through $c_{\alpha/\beta,\vec{k}}(t) = \sum_{p}c_{\alpha/\beta,\vec{k},p}e^{-i\omega p t}$. The Hamiltonian given in Eq.(\ref{eq.maj_ham_timed}) goes to:
\begin{equation}
\label{eq.maj_ham_timed_frequency}
H_{\vec{k}}(t) = \sum_{p,p'}\sum_{\alpha,\beta} c^{\dagger}_{\alpha,\vec{k},p'}(t)h^{\alpha,\beta}_{\vec{k}}(t)c_{\beta,\vec{k},p}(t)e^{i\omega t (p-p')}
\end{equation} 
{Substituting the above relation} in the  Hamiltonian in Eq.(\ref{eq.Ham_floq_bloch}) and {recalling
the definition in  Eq.(\ref{eq.inner_prod})}, we obtain the following inner product, 
\begin{equation}
\label{eq.ham_inner_prod}
\braket{\braket{u_{\alpha, \vec{k}, p'}|\mathcal{H}_{\vec{k}}(t)|u_{\beta, \vec{k}, p}}} = \tilde{h}^{\alpha,\beta}_{p',p} - p\omega\delta_{p,p'}\delta_{\alpha,\beta}
\end{equation}
with 
\begin{equation}
\label{eq.ham_integration}
\tilde{h}^{\alpha,\beta}_{p',p} \equiv \frac{1}{T}\int_{0}^{T}h^{\alpha,\beta}_{p',p}(t)e^{i\omega t (p-p')}dt.
\end{equation}
{Eq.(\ref{eq.ham_inner_prod}) reveals that the static tight binding (TB) model in $N$ dimensions under the application of an AC electric field gets mapped to a time-independent TB model in $(N+1)$ dimensions with an effective static DC electric field, i.e.,
the driving frequency ($\omega$), acting along the added dimension.  This effective electric field breaks the translational symmetry along the additional dimension; we can now separate out two regimes on the basis of high or low effective electric field, i.e, in terms of high or low driving frequency. Of course, one needs to compare it against the other energy scale in our model which is the tunnelling strength or the strength of the exchange interactions ($J_{1,2,3}$). The $N$ dimensional lattice thus repeats itself along the (asymmetrical with respect to frequency) extra dimension hosting lattice points that are linked with their neighboring lattice points via renormalised (or photon dressed) tunneling or exchange interaction strengths.}


\section{Merging of Dirac points}
\label{sec_merge}
To drive the system we choose a half wave rectified elliptically polarised electric field given by:
\begin{figure} 
\centering \includegraphics[height = 5cm, width = 5cm]{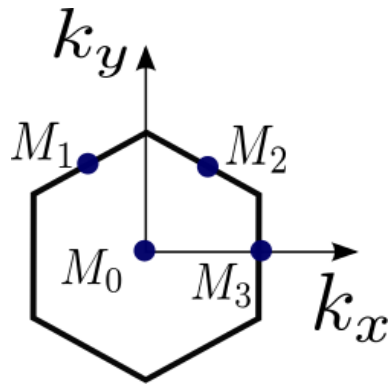}
\caption{The Brillouin Zone (BZ) of the honeycomb lattice is shown, indicating the four TRIM points as has been stated in the text.} 
\label{fig.TRIM}
\end{figure}

\begin{eqnarray}
\label{eq.Efield}
E(t) &&= A(\sin(\omega t), \sin(\omega t + \phi),0),\hspace*{2mm} 0<t<\frac{T}{2} \\ \non
E(t) &&= 0 \hspace*{36mm} \frac{T}{2} \leq t < T
\end{eqnarray}
{where $A$ is the magnitude of the field, the phase $\phi$ is the measure of the ellipticity and the half wave rectified electric field that repeats itself over a time period $T$ with the frequency denoted by $\omega=\frac{2\pi}{T}$. The {motivation behind using} a half wave pulse (and not a complete sinusoid) is to prevent the integral in Eq.~(\ref{eq.ham_inner_prod}) from generating {interactions of $\delta$ function form  rendering the dynamics trivial}.}

  {Although the reason behind choosing a half wave rectified sinusoidal electromagnetic wave is to have an effective renormalised Hamiltonian with competing terms of finite strength,  these waves are indeed feasible experimentally and are commonly generated in optical setups via a Mach Zender optical modulator. This modulator functions by splitting the incoming electromagnetic beam into two waveguides where a time varying voltage in one of the waveguides induces a time-dependent phase shift (via the electro-optic effect) in the passing wave. When the undisturbed wave passing through one waveguide is made to interfere with the phase shifted wave from the other waveguide, the two waves interfere and the time-dependent phase shift manifests itself as an (time varying) amplitude modulation of the initial wave. Thus, the amplitude of a laser beam may be tuned to obtain a half wave rectified sinusoidal electromagnetic wave by suitably varying the applied voltage (periodically) in the setup discussed above.}

{One should now work in the high frequency regime ($\omega\gg J_1,J_2,J_3$) to show that the above mentioned driving scheme can make the Dirac points merge. The high frequency regime as has been previously mentioned, has been identified by comparing the two energy scales in the problem : the driving frequency and the strength of the exchange interactions. Now Eq.~(\ref{eq.ham_inner_prod}) tells us that the Floquet Hamiltonian matrix is infinite dimensional in the space defined by $p,p'$ as they belong to the set of all integers. Thus, at such high frequencies, the last term in Eq.~(\ref{eq.ham_inner_prod}) is more dominant, and hence  the Floquet Hamiltonian matrix is roughly block diagonal. The Floquet sub-bands are hence, uncoupled {which allows us to set   $p \to p'$ in Eq.~(\ref{eq.ham_inner_prod})}. Hence we can restrict ourselves to a system described by a $(2\times 2)$ Hamiltonian similar to Eq.(\ref{eq.kit_timed_bdg_maj}).} 
This leads to a renormalization of the hopping strengths $J_{i}\to J'_{i}$ (Eq.(\ref{eq.ren_hop_strengths})). For our analysis we restrict ourselves to the  block with $p = 0$. This implies that the quasi-energy $\epsilon_{\alpha} = |\tilde{\rho}|$. The system is studied using the spectrum generated for this reduced Hamiltonian {with modified coupling strengths recalling Eq.~(\ref{eq_Pi}) and the form of the electric field  (\ref{eq.Efield}) (see Appendix for details)}

\begin{eqnarray}
\label{eq.ren_hop_strengths}
J'_{1} &&= J_{1} + \frac{A}{\pi}\left(-\sqrt{3} + \cos\phi\right)\Pi_{1} \nonumber \\ 
J'_{2} &&= J_{2} + \frac{A}{\pi}\left(\sqrt{3} + \cos\phi\right)\Pi_{2}  \nonumber \\
J'_{3} &&= J_{3} - \frac{2A}{\pi}\Pi_{3}\cos\phi
\end{eqnarray}

{It is now necessary}  to obtain a condition involving the renormalized parameter $J'_{1},J'_{2}$ and $J'_{3}$ for the merging of Dirac points. Due to time reversal symmetry of the Hamiltonian, the structure of the Dirac points in the $k$ space is such that if there occurs a Dirac point at $\vec{D}$, another  Dirac point must be present at $-\vec{D}$. When two such Dirac points merge, it implies that a point in the $k$-space is mapped to itself under time reversal symmetry, this is called a time reversal invariant momentum (TRIM) point. To obtain the TRIMs for the Kitaev model we note that the two Dirac cones must move equal distances in $k$-space along the lattice vectors $\vec{M}_{1}$ and $\vec{M}_{2}$. Thus giving the TRIM points as $\frac{1}{2}(p\vec{M}_{1}+q\vec{M}_{2})$ with $(p,q) = (1,1),(1,0),(0,1),(0,0)$ (see
Fig.~(\ref{fig.TRIM})). 
This yields the following conditions on the hopping parameters:
\begin{eqnarray}
\label{eq.Dirac_merging}
&&M_{0}\to\hspace*{1mm}J'_{1} + J'_{2} + J'_{3} = 0 \hspace*{2mm} M_{1}\to\hspace*{1mm} J'_{1} = J'_{2} + J'_{3}  \nonumber\\
&&M_{2}\to\hspace*{1mm}J'_{2} = J'_{1} + J'_{3} \hspace*{5mm} M_{3}\to\hspace*{1mm}J'_{3} = J'_{2} + J'_{1}
\end{eqnarray}

\begin{figure} 
\centering \includegraphics[height = 7cm, width = 8cm]{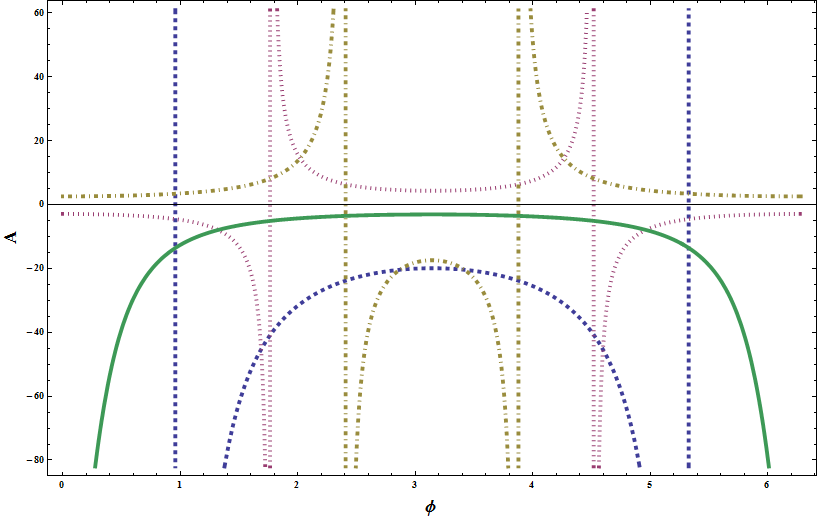}
\caption{(color online) The phase diagram for the anisotropic case $\ga_{1} ~\neq ~ \ga_{2} ~\neq ~ \ga_{3}$. The diagram shows the lines in $(\phi,A)$ space where the Dirac points merge. The position of merge is indicated by the $M_{i}$ labels ($M_{0} =$ blue (dotted) , $M_{1} =$ green (solid) , $M_{2} =$ brown (dot-dashed), $M_{3} =$ pink (dashed)). The points where two such merge lines intersect represent regions where one of the couplings $J'_{i}$ is zero. The phases are distinguished by the colour of the boundary walls as elaborated in the text.}
\label{fig.phase_diag1}
\end{figure}

\begin{figure} 
\centering \includegraphics[height = 7cm, width = 8cm]{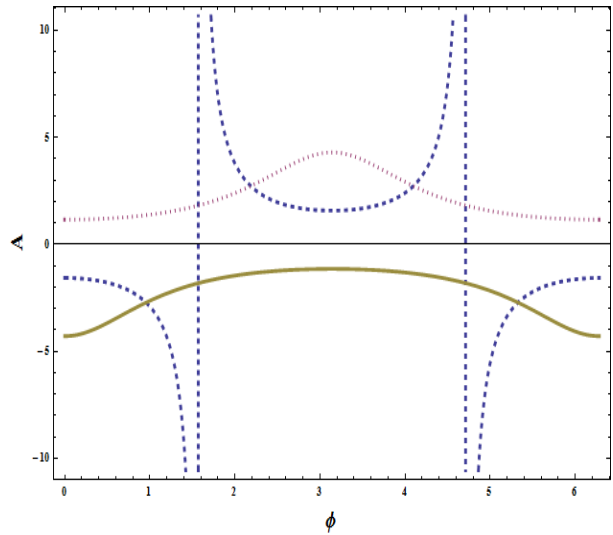}
\caption{(color online) The phase diagram for the isotropic case $\ga_{1} ~=~ \ga_{2} ~=~ \ga_{3}$. The diagram shows the lines in $(\phi,A)$ space where the Dirac points merge. The position of merge is indicated by the $M_{i}$ labels ($M_{1} =$ brown (solid), $M_{2} =$ pink (dotted), $M_{3} =$ blue (dashed)). The condition for merge at $M_{0}$ is missing because of the isotropy. The phases are distinguished by the colour of the boundary walls as elaborated in the text.}
\label{fig.phase_diag2}
\end{figure}

These relations are the same as the ones for phase transition derived for the Kitaev model without the ME coupling (Eq.(\ref{eq.Kit_ham_bdg})) \ct{kitaev06}. The noticeable difference is now the conditions are imposed on the renormalized parameters $J'_{i}$s.
It should also be noted that with the renormalized coupling parameters, it is now possible to merge the Dirac points at the center of the BZ even if $\sum_{i}J_{i}\neq 0$; this is because the merging conditions are imposed on the renormalized strengths $J'_{i}$'s and does no longer on bare coupling strengths $J_{i}$'s. The merging of Dirac points at the TRIM points produces a semi-Dirac kind of a spectrum which means that the spectrum is linear along one direction while being quadratic along the other. We can now vary $A$ and $\phi$ to achieve a transition from a Dirac to a semi-Dirac spectrum.\\

\section{The Phase Diagrams}
\label{sec_phase}

\begin{figure*}[ht]
\centering
\subfigure[~PDPs at $A = 0$, well inside the BZ]{%
\includegraphics[width=0.45\textwidth,height=5cm]{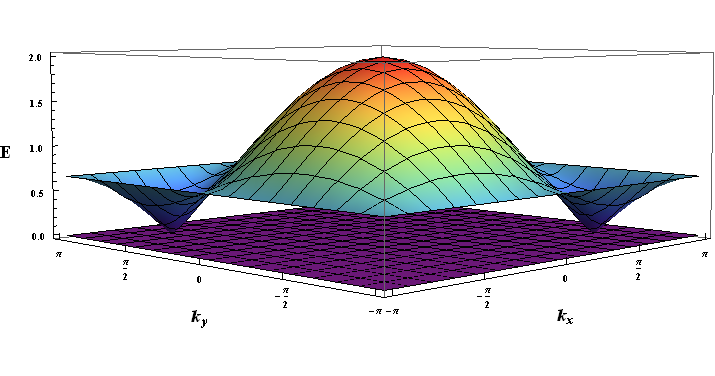}
\label{fig:subfigure1}}
\quad
\subfigure[~The PDPs now lie closer to the BZ boundary at $A = 1.1$]{%
\includegraphics[width=0.45\textwidth,height=5cm]{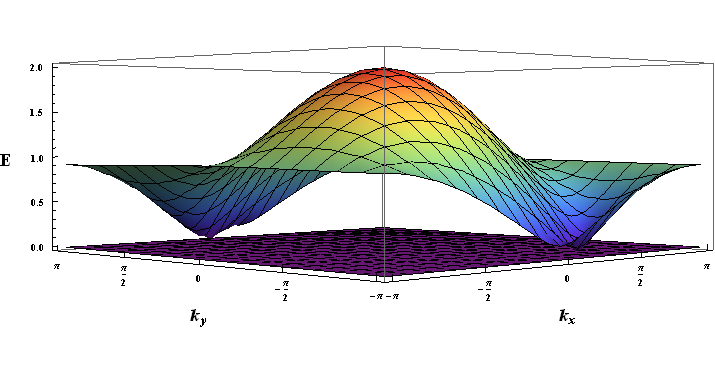}
\label{fig:subfigure2}}
\subfigure[~Semi-Dirac spectrum : The PDPs have now merged at one TRIM point (M2 here) at the toroidal BZ boundary at $A=1.382528373335$]{%
\includegraphics[width=0.45\textwidth,height=5cm]{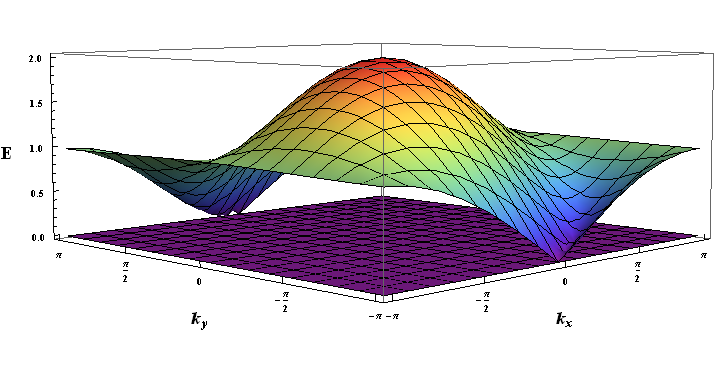}
\label{fig:subfigure3}}
\quad
\subfigure[~Gapped quasi-energy spectrum at $A=2.1$]{%
\includegraphics[width=0.45\textwidth,height=5cm]{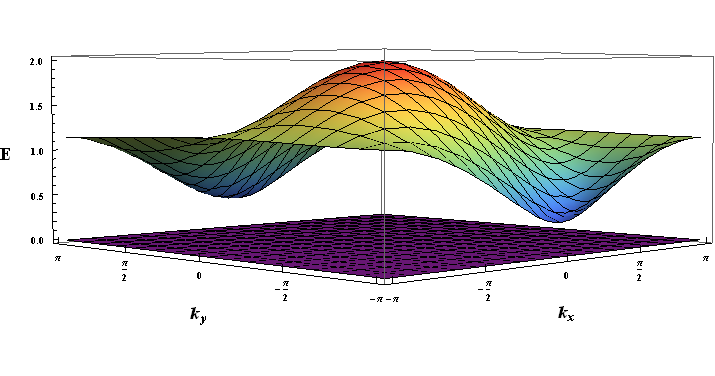}
\label{fig:subfigure4}}

\caption{This k-space quasi-energy spectrum (denoted by $E$ here) shows the dynamic merging of the PDPs in the gapless phase of the undriven Kitaev model for system parameters $J_{1}=J_{2}=J_{3}=\frac{1}{3}$ with a field amplitude varying from (a) $0$ to (b) $1.1$ going up to (c) $1.382528373335$ and then to (d) $2.1$ at a fixed phase of $\phi = 1$ and all $\gamma_i = 0.5$ (isotropic case). We see in (a) that there are 2 conical (anisotropic linear dispersion) PDPs touching the zero quasi-energy plane within the gapless phase of the undriven Kitaev model in the first BZ. As the field strength varies, we notice that although the quasi-energy spectrum is still gapless in (b), the 2 PDPs have shifted towards the BZ boundary. In (c), the amplitude of the field is such that the 2 PDPs now lie at the BZ boundary. But, since the quasi-momentum ($k_x,k_y$) lie on a torus, the two PDPs are actually at the same quasi-momentum (TRIM) point. Hence, the Dirac points have merged to produce a semi-Dirac spectrum in the gapless phase of the undriven Kitaev model under driving at this TRIM point. We further observe in (d) that a slight variation of the field amplitude only manages to gap out the spectrum as expected.}
\label{fig.Dirac_merge_contour}
\end{figure*}

The phase diagram for the merging scenario will depend on all the parameters $J_{i}$'s, $A$, $\phi$ and the $\Pi_{i}$'s where only $A$ and $\phi$ serve as our externally tunable. We will next demonstrate how a pair of Dirac points (PDPs) located at the center of the undriven gapless region in the parameter space of $J_i$'s can be merged through driving to produce a semi-Dirac spectrum. Two kinds of phase diagrams emerge depending on the anisotropies in polarisation. We will gradually discuss them below.\\

To draw the phase diagrams we assume that the EM field affects the underlying lattice in such a way so that the polarizability constants  ($\Pi_{i}$'s) will only be a fraction of the exchange constants ($J_i$'s). Thus, we can write : 
\begin{equation}
\label{eq.polar_renorm}
\Pi_{i}=\gamma_{i} J_{i}
\end{equation}
where $\gamma_{i} < 1$. \\

Let us now discuss the first scenario in which all the $\gamma_i$'s are unequal. The Fig.(\ref{fig.phase_diag1}) shows the different phases that can emerge from such a condition. The Dirac points may merge at a particular TRIM point $M_i$ for different values of amplitude $A$ and $\phi$. Thus, the four continuous lines in four different colors (one color for each $M_i$, $i=0,1,2,3$) in Fig.(\ref{fig.phase_diag1}) are the merging lines where a PDP get annihilated rendering a semi-Dirac spectrum. There are four such lines because there are four TRIM points as has been discussed earlier.\\

We can go from one phase to the other in the phase diagram only upon crossing the merging lines. The phases, as a result, can be identified by observing whether in that phase a PDP can merge at all the four TRIM ($M_i$) points or it can merge at all the TRIM points ($M_i$'s) except one $M_i$ where it can never merge. The phases in which the four different merging lines serve as the phase boundary are phases in which a PDP can merge at any one of the TRIM ($M_i$) points. There are also phases in which only three (or two) different merging lines form the phase boundary. In such phases, a PDP can can never merge at those $M_i$ points (one or two respectively) whose merging lines do not constitute the phase boundary. The phases have thus, accordingly been identified in the figure (Fig.(\ref{fig.phase_diag1})).\\

A secondary route towards merging is suggested by the structure of the renomalized $J$'s (see Eq.(\ref{eq.ren_hop_strengths})), namely the isotropic case where $\ga_{1} = \ga_{2} = \ga_{3}$. The resulting phase diagram is shown in Fig.(\ref{fig.phase_diag2}). The point to be noted in such a case is that the PDPs can merge at two or three TRIM points but not at all four. A pair of PDP's can never merge at the central TRIM point ($M_{0}$) given by the condition $J'_{1} + J'_{2} + J'_{3} = 0$, as is evident from the fact that for the isotropic case $\sum_{i}J'_{i} = \sum_{i}J_{i} \neq 0$.\\

This fact is reflected in the phase diagram. Characterizing the phases using the same classification scheme as used for the earlier anisotropic case; we find the existence of phases for which the boundary walls represent merger of two or three different TRIM points. It can also be seen that the phase represented by low values of the field amplitude $A$ has three different boundary walls so from this phase we can effectively merge the PDP's at all the TRIM points except at $M_{0}$.

The phase diagram of the isotropic case ($\gamma_1=\gamma_2=\gamma_3$) is indeed very simple to understand. A plot of the quasi-energy spectrum ($E$) against the quasi-momentums given by ($k_x,k_y$) as a means to illustrate the isotropic case is shown in figure (Fig.(\ref{fig.Dirac_merge_contour})). We vary the field amplitude $A$ from $0$ to $2.1$ keeping the phase fixed at $\phi=1$. The quasi-energy spectrum changes from having a PDPs to a merger at a TRIM point on the $M_2$ line (pink line in Fig. (\ref{fig.phase_diag2})) yielding a semi-Dirac spectrum at $A = 1.382528373335$. 

  {The Dirac points (DPs) which are time-reversed partner of each other can merge at the time reversal invariant momentum (TRIM) points only. There are seven TRIM points with one of them at the centre and the other six at the zone boundaries of a hexagonal Brillouin zone (BZ). But, since the TRIM points at the zone boundaries are equivalent modulo (or related via) a reciprocal lattice vector, there are only three inequivalent TRIM points at the zone edges of a hexagonal BZ. This takes the total tally of inequivalent TRIM points within a hexagonal BZ to four (three at the BZ boundary and one at the centre).  To make the scenario further transparent,  we refer to the Fig.~(\ref{fig.new}) that explicitly tracks how the DPs undergo a merging transition as the vector potential (A) is varied. In the figure, we depict how a merging transition happens when one DP moves and settles at one of the TRIM points while its time reversed partner settles at another TRIM point, where both the TRIM points are related by a reciprocal lattice vector. Here, one must note that the merging of DPs does not happen at the centre of the BZ; rather  
it occurs  at two points at opposite edges  related via a reciprocal lattice vector and hence are equivalent.}

\begin{figure} 
\centering \includegraphics[height = 7cm, width = 8cm]{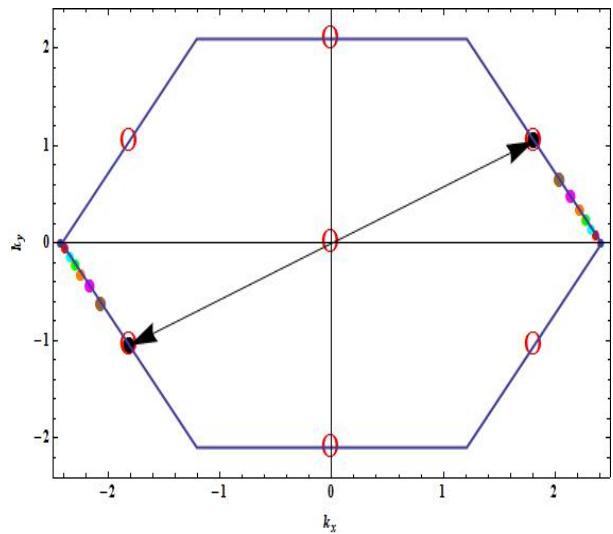}
\caption{(color online)   { The BZ boundary has been plotted in blue while the TRIM points lying on it  are shown in hollow red dots. We mark the DPs  as little solid dots of varying colors. There are only two DPs for a fixed vector potential strength (A) and hence, we see pairs of DPs in the same colour  (and size) on the two opposite edges of the BZ.  The growing size of the dots indicate the direction in which the DPs proceed nearly along the BZ boundary towards the TRIM points before merging when $A$ has been varied from $0$ (DPs in small blue at the BZ corner which lies on the $k_x$ axis) to $0.2$ (DPs in big red dot) to $0.4$ (DPs in bigger cyan) to $0.6$ (DPs in even bigger green) to $0.8$ (DPs in large orange) to $1.0$ (DPs in large magenta) to $1.2$ (DPs in larger brown)  to finally at $A = 1.382528373335$.   The DPs merge for
$A= 1.382528373335$ as they lie on top of the red TRIM point (black solid huge dot inside). The black arrow depicts that the two TRIM points in which the DPs merge differ only via a reciprocal lattice vector. 
 }}
\label{fig.new}
\end{figure}

 A similar plot for the anisotropic case ($\gamma_1\neq\gamma_2\neq\gamma_3$) can also be easily constructed. A careful analysis then would clearly corroborate the fact that the anisotropic case only differs from the isotropic case in the presence of the possibility that the PDPs can also come together at the $M_0$ TRIM points.    

\section{Conclusion}
\label{sec_conc}
In summary, through this work it is clear that external controls can be used to effectively move a PDP in the reciprocal space for the Kitaev honeycomb Hamiltonian. In our analysis we have used a modified form of the Hamiltonian with an effective magneto-electric coupling to facilitate our manipulation of the location of the PDP's through an externally applied half wave rectified electric field. The effective phase diagram of the system is then critically dependent on the amplitude $A$ and phase $\phi$ of the external field.\\
The important point is that through the scheme outlined the phases of the model can be externally controlled and indeed fine tuned phases can be created where the model reduces to an effective Ising model (at the intersection of two merge lines).
  {Finally, we note that merging of DPs involves transition from gappless to gapped phases which may
be experimentally captured  in the entanglement entropy \ct{islam15} as also theoretically predicted in \ct{mandal16}.}

\begin{acknowledgments}
We acknowledge  Diptiman Sen for critical comments on our work.  AD acknowledges SERB,  DST, India,  for financial support. 
\end{acknowledgments}

\appendix*
\section{The renormalised coupling constants}

In this appendix, we briefly discuss how the renormalised coupling constants given by $J'_i$s in Eq.~(\ref{eq.ren_hop_strengths}) can be obtained. The main aim of this calculation is to get rid of the time dependence in the Hamiltonian using the periodic nature of the driving by diagonalising the Hamiltonian in the Sambe space using the composed inner product as given in Eq.~(\ref{eq.ham_inner_prod}).The further use of Eqs.~(\ref{eq.timed.breakup.ham}) and (\ref{eq.ham_integration} to perform the composed inner product yields :
\begin{equation}
\label{eq.appendix_integration}
\tilde{h}^{\alpha,\beta}_{p',p}=\frac{1}{T}\int_{0}^{T}\left(A_{\vec k}(t)-iB_{\vec k}(t)\right)e^{i\omega t (p-p')}dt
\end{equation}
where each of the $\tilde{h}^{\alpha,\beta}_{p',p}$ can be cast as a $2\times 2$ matrix of the form :
\begin{equation}
\label{eq.appendix_ham_freq}
H_{p'-p} = \left(\begin{array}{cc} 0 & \rho_{p'-p}(\vec{k})  \\ \cr \rho_{p'-p}^{*}(\vec{k}) & 0 \end{array} \right)
\end{equation}
and $\rho_{p'-p}(\vec k) = \tilde{h}^{1,2}_{p',p}$\\

Since our interest lies in the high frequency regime where the Floquet sub bands decouple and form $2\times 2$ blocks for each $p$, the limit $p'\rightarrow p$ is taken in the above Eq.~(\ref{eq.appendix_integration}) alongwith the choice of the $p=0$ block as has been mentioned earlier, to obtain $\rho_0(\vec k) = A'_0(\vec k)-iB'_0(\vec k)$, where $A'_0(\vec k)$ and $B'_0(\vec k)$ are :
\begin{widetext}
\begin{eqnarray}
\label{eq.appendix.A0kB0k}
A'_0(\vec{k}) &&= 2\left[\left\{J_{1} + \frac{A}{\pi}\left(-\sqrt{3} + \cos\phi\right)\Pi_{1}\right\}\sin(\vec{k}.\vec{M_{1}}) - \left\{J_{2} + \frac{A}{\pi}\left(\sqrt{3} + \cos\phi\right)\Pi_{2}\right\}\sin(\vec{k}.\vec{M_{2}})\right] \\ \non
B'_0(\vec{k}) &&= 2\left[\left\{J_{3} - \frac{2A}{\pi}\Pi_{3}\cos\phi\right\} + \left\{J_{1} + \frac{A}{\pi}\left(-\sqrt{3} + \cos\phi\right)\Pi_{1}\right\}\cos(\vec{k}.\vec{M_{1}}) + \left\{J_{2} + \frac{A}{\pi}\left(\sqrt{3} + \cos\phi\right)\Pi_{2}\right\}\cos(\vec{k}.\vec{M_{2}})\right]
\end{eqnarray}
\end{widetext}
Thus, it can easily be seen that in the high frequency regime, when our decoupled Floquet sub blocks have the same form as our  original Kitaev Hamiltonian (see Eq.~(\ref{eq.timed.breakup.ham})), the terms within the curly braces are nothing but the renormalised coupling constants $J'_i$s as has been provided in Eq.~(\ref{eq.ren_hop_strengths}).

\end{document}